\documentclass{aa}  
\usepackage{hyperref}
\usepackage{graphicx}
\usepackage{txfonts}

\newcommand{\nicer}{{\it NICER\/}} %

\newcommand{\swift}{{\it Swift\/}}
\newcommand{\maxi}{{\it MAXI\/}} %

\newcommand{\hxmt}{{{\it Insight}-HXMT\/}}

\def\chiq{$\chi^2$}

\def\be{\begin{equation}} 
\def\ee{\end{equation}}

\begin{document}

   \title{Type I\ X-ray bursts' spectra and fuel composition from the atoll and transient source 4U 1730--22}

   \subtitle{}

\author{Yongqi Lu\inst{1}
        \and
        Zhaosheng Li\inst{1}
        \thanks{Corresponding author}
       \and
    Yuanyue Pan\inst{1}
       \and 
       Wenhui Yu\inst{1}
             \and
    Yupeng Chen\inst{2}
    \and
    Long Ji\inst{3}
    \and
    Mingyu Ge\inst{2}
    \and
    Shu Zhang\inst{2}
    \and
    Jinlu Qu\inst{2}
    \and
    Liming Song\inst{2}
    \and
    Maurizio Falanga \inst{4,5}
          }
   \offprints{Z. Li}

   \institute{Key Laboratory of Stars and Interstellar Medium, Xiangtan University, Xiangtan 411105, Hunan, P.R. China\\ \email{lizhaosheng@xtu.edu.cn, panyy@xtu.edu.cn}
               \and
Key Laboratory of Particle Astrophysics, Institute of High Energy Physics, Chinese Academy of Sciences, 19B Yuquan Road, Beijing 100049, China
\and
School of Physics and Astronomy, Sun Yat-Sen University, Zhuhai, 519082,China
               \and
International Space Science Institute (ISSI), Hallerstrasse 6, 3012 Bern, Switzerland
           \and
Physikalisches Institut, University of Bern, Sidlerstrasse 5, 3012 Bern, Switzerland
              }

   \date{Received XX; accepted XX}

  \abstract{\nicer\ observed two outbursts from the neutron star low-mass X-ray binary 4U~1730--22 in 2021 and 2022, which showed a similar spectral evolution in the hardness-intensity diagram. Seventeen type I X-ray bursts were identified in both outbursts. The X-ray burst spectra showed clear deviations from the blackbody model, firstly $\sim10$ s after onset. Adding the enhanced persistent emission due to the Poynting-Robertson drag or the reflection from the accretion disk both significantly improved the fitting results. We found that 12 out of 17 X-ray bursts showed the photospheric radius expansion (PRE) characteristic. Considering the nine PRE bursts out of ten X-ray bursts observed by \hxmt, 78\% of bursts from 4U~1730--22 exhibited PRE. According to the burst rise time, the duration, the local accretion rate, and the burst fuel composition estimated from recurrence time, we propose that these PRE bursts were powered by pure helium. From the touchdown flux of PRE bursts, we estimate the source distance of $ d=7.54\pm{0.46} (X=0)$ kpc for a canonical neutron star with $M_{\rm NS}=1.4M_\odot$ and $R_{\rm NS}=10~{\rm km}$. }

   \maketitle

\section{Introduction}
\label{Sec:intro}
Neutron stars (NSs) in low-mass X-ray binary (LMXB) systems accrete matter from a Roche-lobe overflowing low-mass companion star \citep[M $\lesssim$ 1 M$_\odot$;][]{Frank92}. The accumulated matter on the NS surface can trigger unstable thermonuclear bursts, also known as type I X-ray bursts \citep[see e.g.,][for reviews]{Lewin93,Strohmayer06,Galloway21}. Based on the composition of the accreted fuel and the ignition depth, most of the observed type I X-ray bursts are burning mixed hydrogen and helium or pure helium. At a low local mass accretion rate, that is $\dot m \lesssim 10\%$ of the local Eddington rate, a helium burst is triggered rapidly and energetically through a triple-alpha process, which only takes an $\sim1-2$ s rise time to reach the peak flux and lasts $\sim 10-20$ s, where helium comes from  either the donor, that is a helium white dwarf, or from the stable completion burning of hydrogen. The decay time of these bursts is attributed to the cooling of the NS photosphere resulting in a gradual softening of the burst spectrum. For a NS LMXB with a higher mass accretion rate, it usually occurs in mixed hydrogen and helium bursts triggered by thermally unstable helium ignition, where hydrogen is accreted faster than it is consumed through a hot CNO cycle limited by $\beta$ decays, then the rise and decay are slower.  For a review, readers can refer to \citet{Lewin93} and \citet{Strohmayer06}, for example.

The X-ray burst spectra can be described by blackbody radiation with temperatures in the range of $kT_{\rm bb}\approx0.5-3$ keV. For some energetic bursts, when the peak luminosity reaches the Eddington luminosity, their radiation pressure exceeds the gravitational potential on the NS surface, causing substantial photospheric radius expansion \citep[PRE;][]{Lewin93}. These PRE bursts are generally used as standard candles to phenomenologically determine the distance to the source \citep{Kuulkers03}. The touchdown moment and the cooling tail during PRE bursts have been used to constrain the NS mass-radius relation \citep{Sztajno87, Ozel09,Poutanen14,Suleimanov17,Li15,Li18}.  From the X-ray burster MINBAR catalog, about 20\% of 
$\sim7000$ observed bursts exhibit PRE bursts \citep{Galloway20}. In a few X-ray bursts, such as 4U 1820--30, 4U 1916--053, and 4U 1728--34, a large fraction of X-ray bursts showed PRE, believed to be powered by pure helium in a hydrogen-poor environment \citep[see e.g.,][]{Galloway08}.

The X-ray source 4U 1730--22 was discovered in 1972 by \textit{Uhuru} during an X-ray outburst \citep{Forman78,Cominsky78}. Based on its spectral evolution and quiescent spectrum, 4U 1730--22 was classified as a possible NS LMXB \citep{Tanaka96,Chen97,Tomsick07}. After an $\sim50$ year-long quiescence state, in 2021 and 2022, 4U 1730--22 went into an $\sim90$ day and $\sim 280$ day outburst, respectively. These two outbursts have been observed by the \maxi/Gas Slit Camera (GSC), \swift,  the Neutron star Interior Composition Explorer \citep[\nicer;][]{Gendreau17}, and \hxmt\ \citep{ATel14683,ATel14686,ATel14688,ATel14757}. 
 \citet{ATel14769} reported the discovery of a type I X-ray burst from 4U 1730--22, confirming its nature as a NS LMXB. \nicer\ observed 17 type I X-ray bursts from 4U 1730--22 for both outbursts, and one of them showed a burst oscillation, confirming that the source is a transient LMXB, hosting a fast rotational NS with a spin frequency of $\sim 585$ Hz \citep{Li22}. \citet{2022arXiv220910721C} reported ten type I X-ray bursts from 4U 1730--22 observed by \hxmt, and they found that nine of them exhibited PRE. The burst spectra of five bright PRE bursts showed  deviations from a blackbody below 3 keV and above 10 keV, which can be modeled either by enhanced persistent emission due to the Poynting-Robertson drag \citep{Walker92,Zand13,Worpel13} or through Comptonization of the burst emission by the corona. From the PRE  bursts,  they derived a source distance of $\sim$10.4 kpc, assuming peak luminosity at the Eddington value for a pure helium burst.  The optical counterpart of 4U 1730--22 emits strong hydrogen emission lines with relatively low full-width half-maximum and weaker helium lines,  suggesting that this source has a main sequence companion, a short orbit period of $\lesssim2-3$ hr, and low inclination with respect to the line of sight \citep{ATel14693,ATel14694}.

In this work, we report the analysis of the 17 X-ray bursts 4U~1730--22 observed by \nicer\ during its 2021 and 2022 outbursts. In Section \ref{Sec:observe}, we describe the properties of the 2021 and 2022 outbursts and X-ray bursts, respectively. In Section \ref{sec:spec_analysis}, we analyze the persistent emission and time-resolved spectra for the X-ray bursts. In Section \ref{Sec:discussion}, we estimate the source distance and the burst fuel composition.

\section{Observations}
\label{Sec:observe}

We analyzed all of the 4U 1730--22 \nicer\ observations between modified Julian date (MJD) 59374--59455  for the 2021 outburst and  between MJD 59624--59860 for the 2022 outburst, respectively. The ObsIDs include 4202200101--420220010134 for the 2021 outburst and 420220010134--420220010143, 5202200101--5202200122, and 4639010101--4639010215 for the 2022 outburst, which have the net unfiltered exposure time of 102 ks and 588 ks, respectively. We processed the \nicer\ data by applying the standard filtering criteria using HEASOFT V6.30.1 and the \nicer\ Data Analysis Software (NICERDAS). 

We extracted 1-s binned light curves in the energy ranges between 0.5--10 keV, 2.0--3.8 keV, and 3.8--6.8 keV using the command ${\tt xselect}$. We also calculated the hardness ratio between 2.0--3.8 keV and 3.8--6.8 keV. From the 2021 and 2022 outbursts, none were found to dip or eclipse. As published by \citet{Li22}, we identified 17 type I X-ray bursts, one during the 2021 outburst and 16 during the 2022 outburst. 

\subsection{The 2021 and 2022 outbursts}
\label{sec:outburst}
\begin{figure}
    \sidecaption
        \includegraphics[width=9cm]{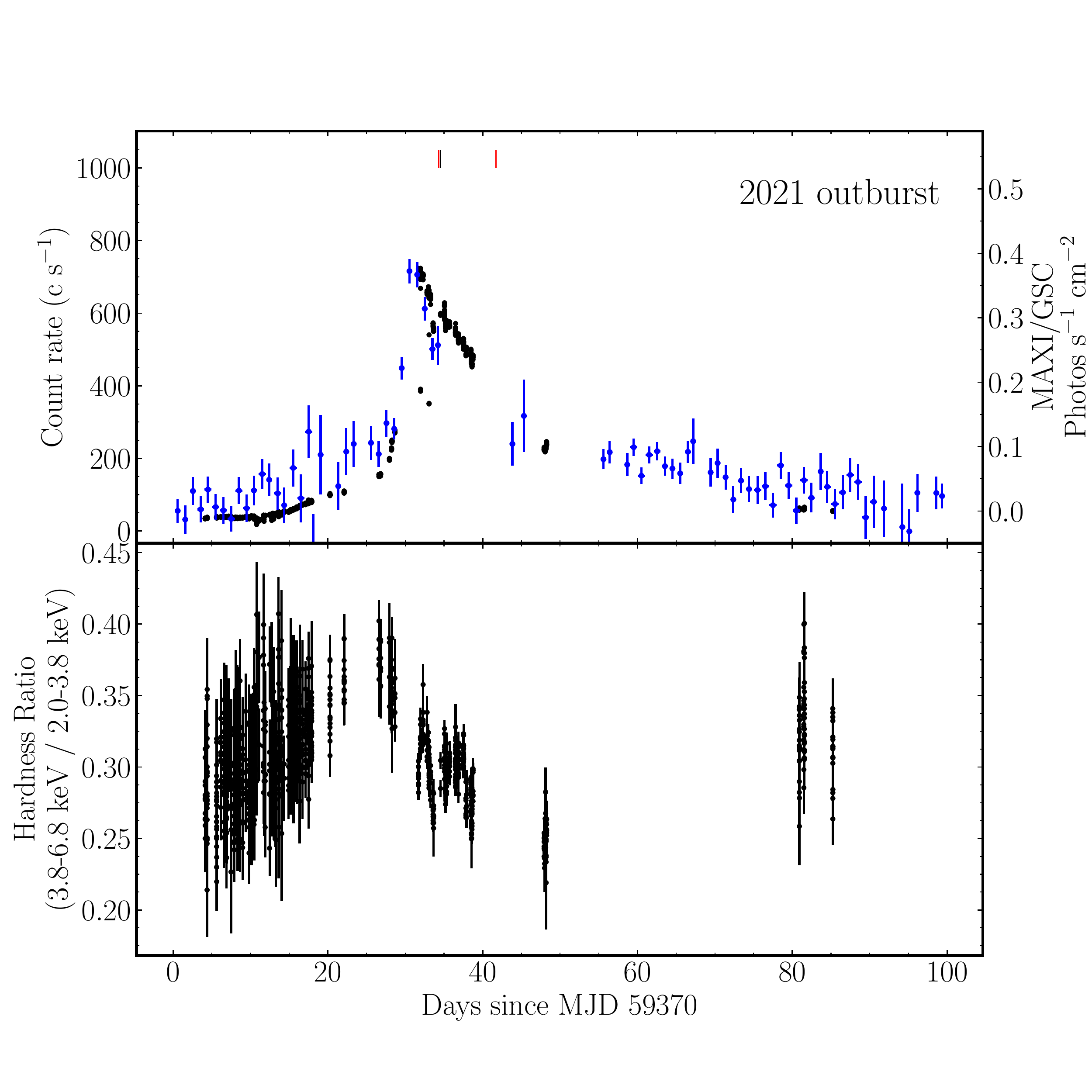}
        \includegraphics[width=9cm]{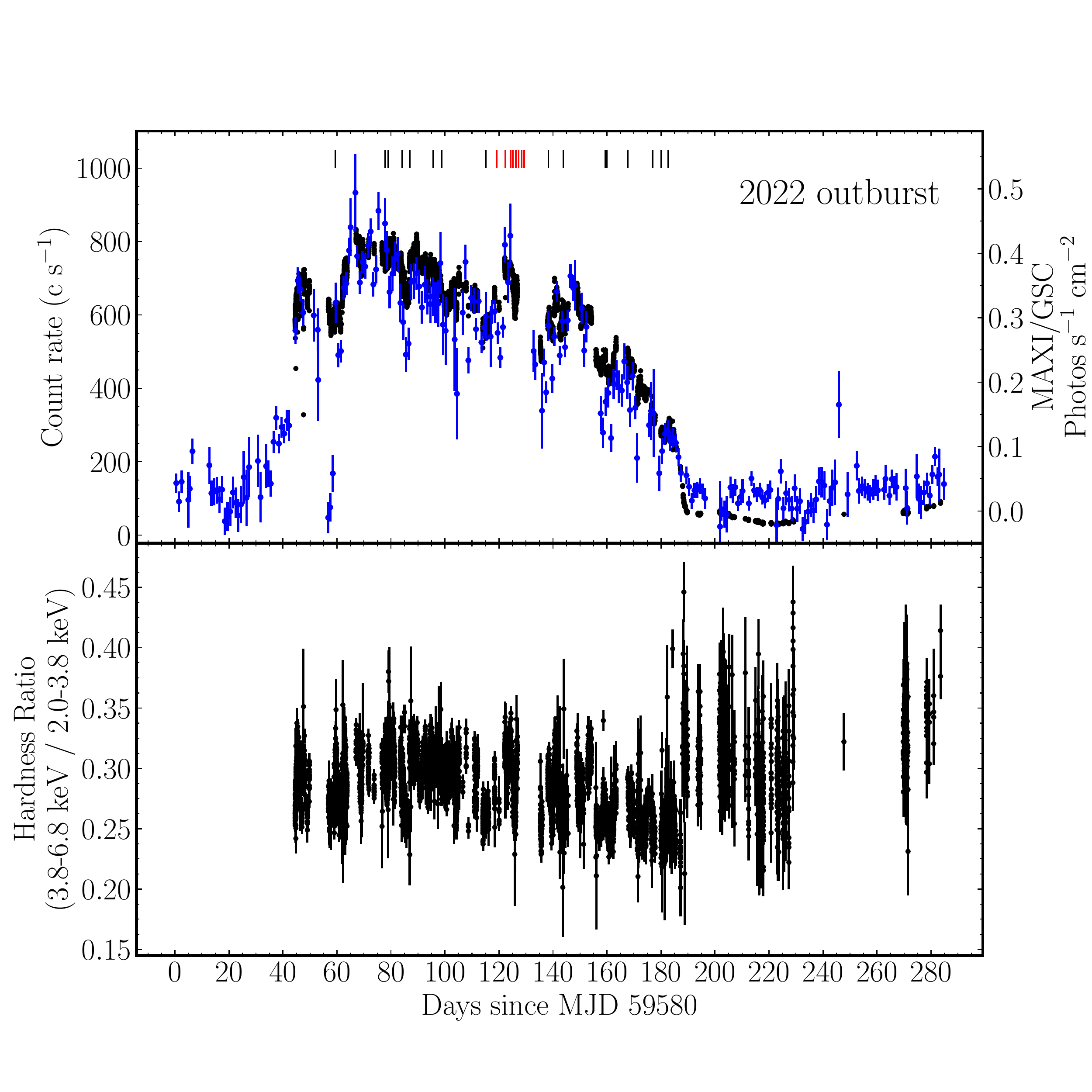} 
\caption{Light curves and hardness ratio of 2021 ({\it top panel}) and 2022 ({\it bottom panel}) outbursts. In the top of each panel, we show the light curves in the 0.5--10 keV and 2.0--20.0 keV bands from \nicer\ and \maxi\ observations, respectively. In the bottom, we show the \nicer\ hardness ratio between 3.8--6.8 keV and 2.0-3.8 keV. Each black point means 64 s data of \nicer\, and all bursts have been removed. The blue points represent the \maxi\ data. We marked the onset of all bursts observed by \nicer\ and  \hxmt\   as black and red bars, respectively.}
\label{Fig:lc}
\end{figure}

In Fig.~\ref{Fig:lc} the 2021 and 2022 outbursts' light curves and hardness ratio are shown. All X-ray bursts have been removed from the light curve; however, the time of the detected bursts are indicated with arrows. To have a better coverage of the two outbursts, that is to say from the onset back to quiescence, we added the \maxi/GCS light curves  in Fig.~\ref{Fig:lc}. These  \maxi/GCS light curves are based on 1.0 d binned monitoring observations in the  2.0-20 keV energy band. 

In the 2021 outburst, during the first 25 days, the \nicer\  source-count rate slowly increased from $\sim40$ ${\rm c~s^{-1}}$ to $\sim200$ ${\rm c~s^{-1}}$; afterwards, the flux rapidly raised within five days to $\sim700$ ${\rm c~s^{-1}}$. After the  outburst maximum, the count rate  decreased within the next two months to a preoutburst flux level of $\sim 50$ ${\rm c~s^{-1}}$.

\begin{figure}
\sidecaption
    \includegraphics[width=9cm]{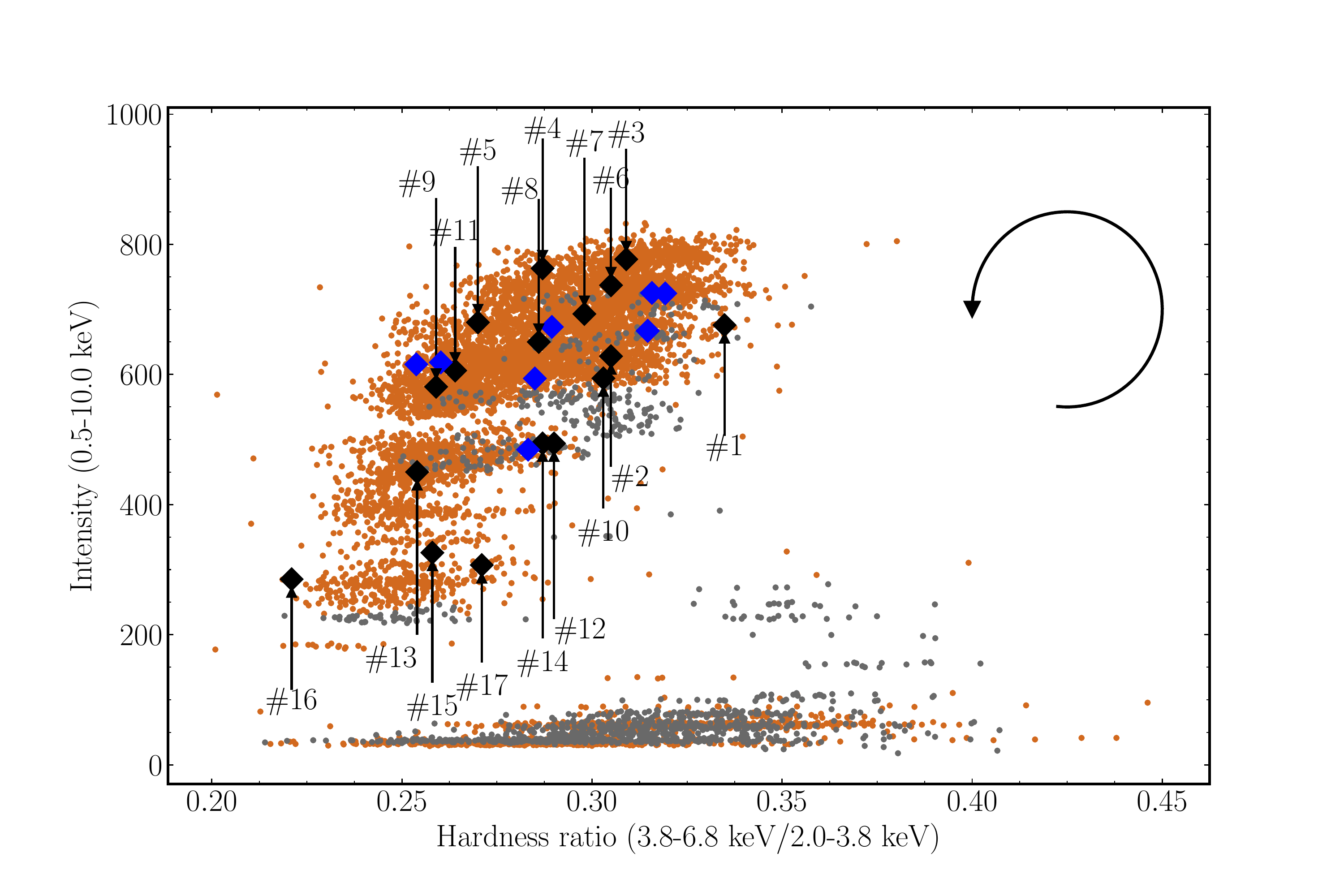}
\caption{
HID of the 4U 1730--22 2021 and 2022 outbursts from \nicer\ observations. The gray (brown) dots represent the outburst observed in 2021 (2022). All bursts have been removed, and each point represents a segment of 64 s. The HID of the persistent emission before each \nicer\ X-ray burst are marked as black diamond points. The blue diamond points represent the \hxmt\ bursts. The circle indicates the evolution direction of HID over time. }
\label{Fig:HID}%
\end{figure}

We found that the hardness ratio increased slowly from 0.25 to 0.35 during the first 25 days and, at the highest flux level, the hardness ratio decreased back to 0.25 to slightly rise again around 0.3 toward the end of the outburst (see Fig.~\ref{Fig:lc}). This is typical behavior of a LMXB atoll source in which the hardness ratio probably evolves as a function of the mass accretion rate \citep{Hasinger89}.
The bottom panel of Fig.~\ref{Fig:lc} shows the 2022 outburst observed by \nicer\ between MJD 59620--59860. The raise of the outburst was missed by \nicer,\ but observed by the \maxi/GSC monitoring program and reported in Fig.~\ref{Fig:lc}. Similar to the 2021 outburst, the  flux raised rapidly to around  $\sim 800$ $\mathrm{c~s^{-1}}$; afterwards, the flux fluctuated for $\sim 100$ days around $\sim 700$ $\mathrm{c~s^{-1}}$, and then it slowly decreased to its preoutburst flux level. The hardness ratio 
followed the same pattern as the 2021 outburst. Indeed in Fig.~\ref{Fig:HID} we show the hardness-intensity diagram (HID) of the 2021 and 2022 outburst from 4U 1730-022, following a spectral transition from a soft to hard state. The X-ray persistent fluxes followed a hardness ration correlation (spectral evolution) similar to other NS LMXB binary atoll sources \citep{Falanga06,Bult18,Zhao22}. These two outbursts are quite similar in their hardness ratio and the HID, even though their durations are different. Moreover, the HIDs of two outbursts generally evolve in a counterclockwise direction. We note that the arrows mark the intensities and the hardness ratios prior to each X-ray burst (see Sec.~\ref{Sec:burst_lc}). We also put the \hxmt\ bursts in the HID, where the intensity and hardness ratio are taken from the \nicer\ data closest to the \hxmt\ burst onset.

\subsection{The X-ray bursts' light curves}
\label{Sec:burst_lc}

\begin{figure*}
\centering
\includegraphics[width=18cm]{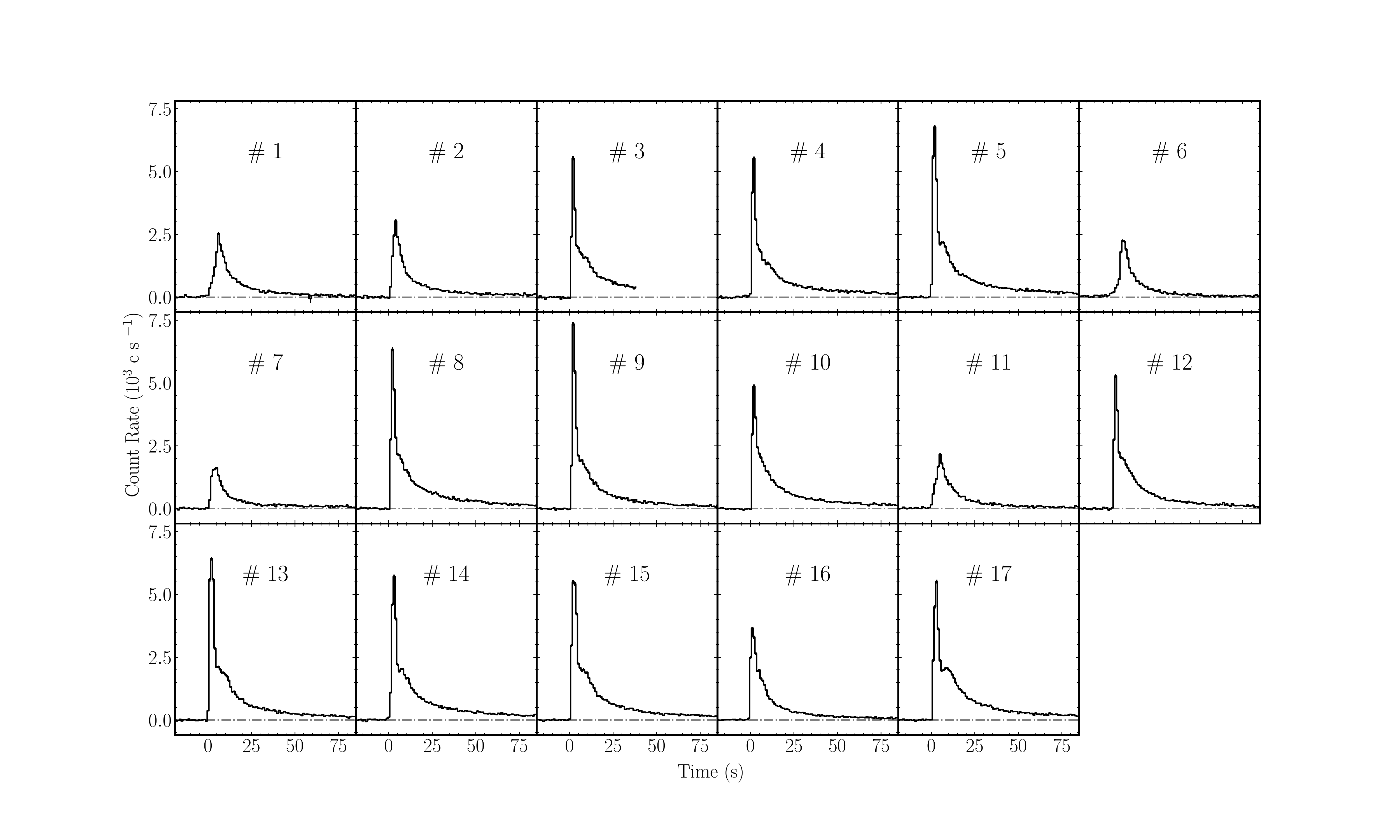}
\caption{Light curve of the 17 X-ray bursts from 4U 1730--22 observed with \nicer. Black lines are light curves in 0.5-10 keV  and they were binned at 1 s time resolution. Light curves were relative to the burst onset time and subtracted the persistent emission. The gray dashed-dotted line represents the persistent emission.}
\label{fig:burst_lc}
\end{figure*}

For each X-ray burst, we first determined the burst peak count rate at the time where the count rate exceeded the preburst rate by a factor of 1.5. We define this time as the onset time of the X-ray burst. The preburst rate is defined as an average count rate in a 64 s window and 10 s prior to the burst peak rate. We determined the burst rise time as the time interval between the burst onset time and the first time bin whose count rates exceed 90\% of the peak count rate. We note that to measure the burst rise time, we used 0.125 s binned light curves instead of a 1 s bin. The burst end time was calculated at the point where the burst count rate decayed to the preburst level. Specifically, we searched the first time bin whose count rate was within $1\sigma$ of the preburst rate after the onset time of 20 s. In Fig. \ref{fig:burst_lc}, we show the net 0.5-10 keV 1--s light curves of all X-ray bursts relative to their onset time. We note that part of the tail in burst \#3 was truncated due to the \nicer\ data gap. The source persistent emission is considered as the background and subtracted from the burst count rate.  From the burst spectral analysis, we could distinguish the non-PRE and PRE bursts as is detailed in Sec \ref{Sec:spec_fa} and Sec \ref{Sec:spec_diskref}, respectively. Bursts \#1--2, \#6--7, and \#11 are non-PRE bursts, with a 3.5--6 s rise time to a burst peak count rate of about 3000 $\mathrm{c~s}^{-1}$. The PRE bursts  have a shorter rise time, $\sim0.5-2$ s, and a higher peak rate of $\sim 6000-8000$ $\mathrm{c~s}^{-1}$. The rise time, peak rate, and observed recurrence time of all bursts are listed in Table~\ref{table:burst_ob}. Burst \#2 was the first burst to be observed during the 2022 outburst, and therefore it did not experience recurrence. We note that, due to the relatively short exposures of each \nicer\ observation, we did not observe successive bursts in a single continuous exposure; therefore, to have the most accurate recurrence time, we also report the ten burst start time observed by \hxmt\ in Table~\ref{table:burst_ob} \citep{2022arXiv220910721C}.

\section{Spectral analysis}   
\label{sec:spec_analysis}
For this work, we performed the spectra analysis using Xspec v12.12.0 \citep{Arnaud96}. The instrumental background spectra were extracted from the tool $\texttt{nibackgen3C50}$ \citep[][]{Remillard21}. We generated the ancillary response files ($\texttt{ARFs}$) with the tool $\texttt{nicerarf}$ and response matrix files ($\texttt{RMFs}$) with the tool $\texttt{nicerrmf}$. The errors of all parameters are quoted at the $1\sigma$ confidence level. 
\subsection{Preburst persistent emission}
\label{Sec:spec_per}
\begin{figure}
\centering
\includegraphics[width=6cm,angle=-90]{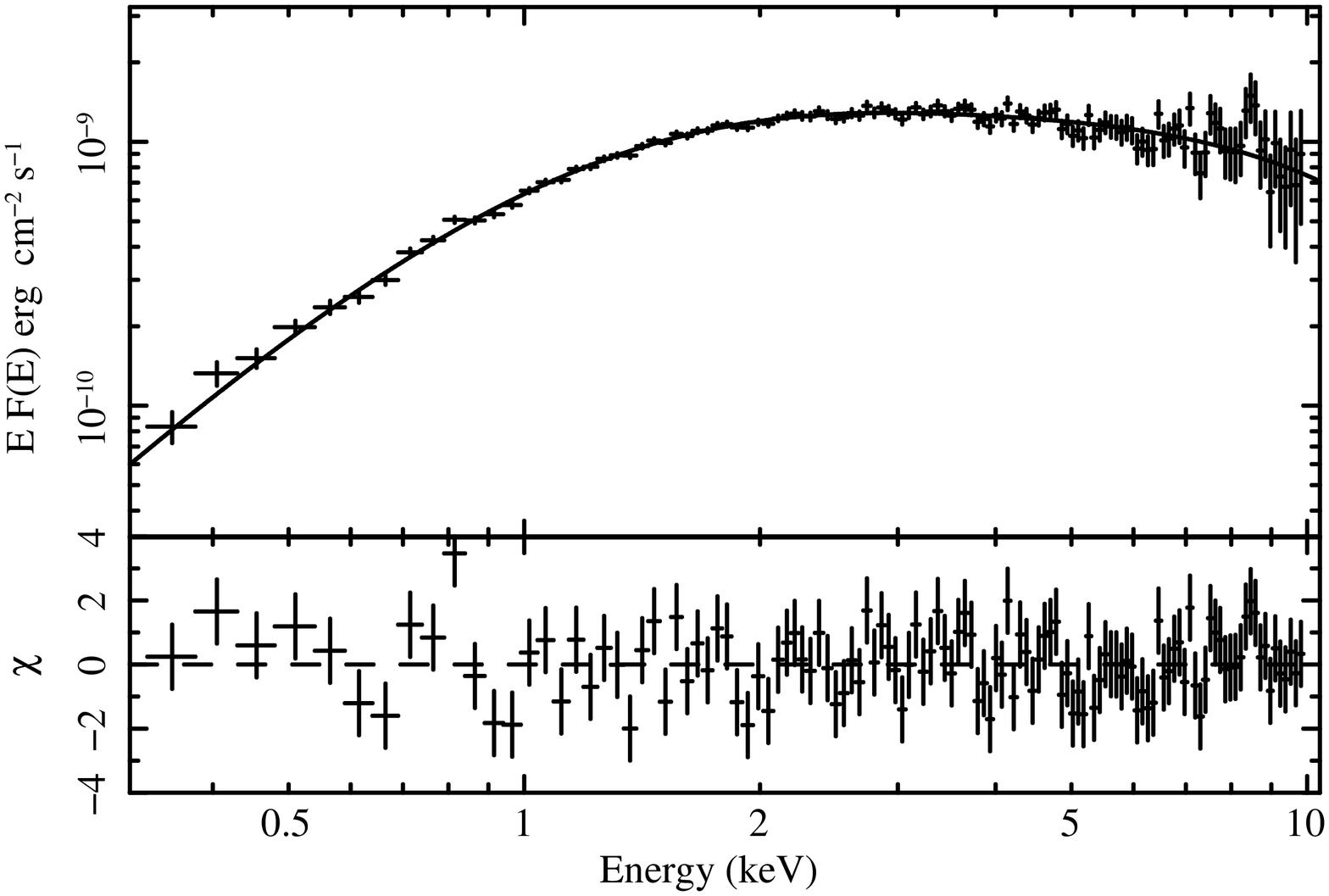}

\includegraphics[width=6cm,angle=-90]{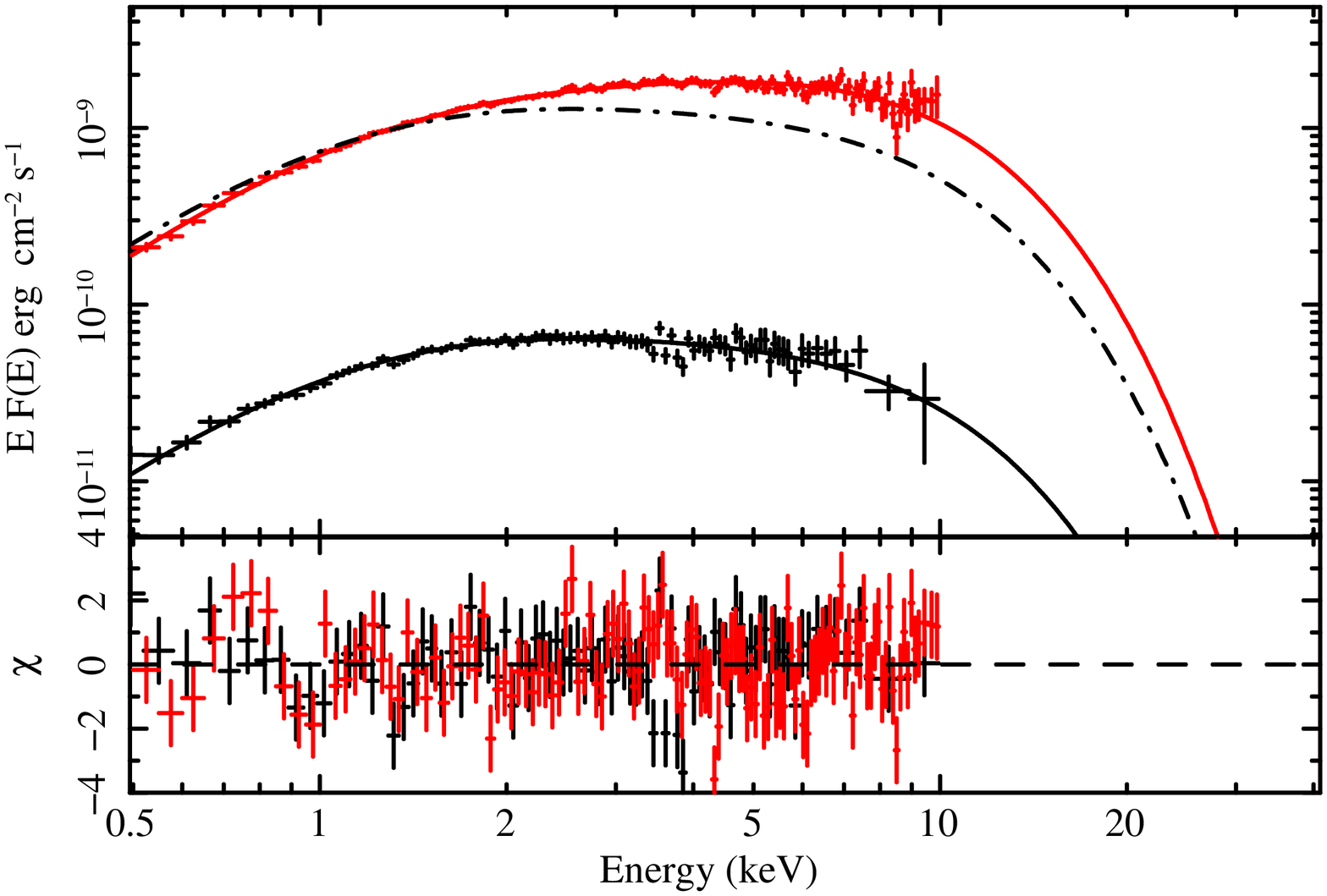}
\caption{Persistent spectra of 4U 1730--22. Top panel: Preburst persistent spectrum in the energy range 0.3--10 keV and the best-fit models, ${\tt Tbabs} \times {\tt nthcomp}$, for burst \#5. Bottom panel: Red points are the persistent spectrum produced from the observations on top right of HID with the exposure of 256 s, while the black points are from the bottom of HID with the exposure of 800 s. The solid lines are the best-fit models for each spectrum. The same models were used. The black dash-dotted line is the black solid model multiplied by 20.  
The residuals of the best-fit model to the data are plotted.}
\label{fig:per}
\end{figure}

For all 17 observed X-ray bursts in 4U 1730--22, we extracted a 128 s long preburst persistent emission spectra. We performed the optimal binning for each persistent spectrum by using {\tt ftgrouppha} as suggested by the \nicer\ team. Then we adopted the thermally Comptonized continuum model, ${\tt nthcomp}$ \citep{Zdziarski96,Zycki99}, modified by the Tübingen–Boulder model, ${\tt TBabs}$, with abundances from \citet{Wilms_2000} to fit all persistent spectra. The model parameters include the asymptotic power law index, $\Gamma$, the electron temperature, $kT_{\rm e}$, the seed photon temperature, $kT_{\rm bb,~seed}$, the input type of seed photons, and the normalization for {\tt nthcomp}, as well as the equivalent hydrogen column, $N_{\rm H}$, for {\tt TBabs}. %
We applied a joint fit to all persistent spectra, where we tied the absorption column density across all spectra, but let the parameters of the {\tt nthcomp} model vary for each spectrum \citep[see e.g.,][]{2021ApJ_Bult}. 
The red $\chi^2$ are around unity, indicating that the persistent spectra are well fitted by the {\tt nthcomp} model. We found the absorption column density at $N_{\rm H} = (0.413 \pm{0.004}) \times 10^{22}~ {\rm cm}^{-2}$, which is well consistent with the value $N_{\rm H} = 0.37_{-0.12}^{+0.07} \times 10^{22} ~{\rm cm}^{-2}$ from the {\it Chandra} observations \citep{Tomsick07}. The prebursts' persistent spectra showed similar parameters, that is the seed photon temperatures, $kT_{\rm bb,~seed}\sim0.36$ keV, and the power law index, $\Gamma\sim1.8-2.2$. We also calculated the unabsorbed bolometric flux in the energy range of 0.1--250 keV by using the tool \texttt{cflux}. The best-fit parameters are listed in Table \ref{table:preburst} and, in the top panel in Fig. \ref{fig:per}, we show one best-fit preburst persistent emission spectrum (burst \#5). We also generated two persistent spectra from the top right -- with the rate $\sim816~\rm{c~s^{-1}}$ and hardness ratio $\sim0.32$ -- and from the bottom left -- with the rate $\sim38~\rm{c~s^{-1}}$ and hardness ratio $\sim0.27$ -- part in HID, with the total exposure of 256 s and 800 s, respectively. The above-mentioned model can also fit the spectra very well by using the same $N_{\rm H}$ (see the bottom panel in Fig.~\ref{fig:per}). The black dash-dotted line, which is the black solid model multiplied by 20, is slightly softer than the red solid line, indicating the hardness ratio difference between these two spectra. 

\begin{table*}
\begin{center} 
\renewcommand\arraystretch{1.2}
\caption{ Best-fitted parameters of all persistent spectra.  \label{table:preburst}}
\resizebox{\linewidth}{!}{\begin{tabular}{ccccccccc} 
\hline\\ %
{\centering \nicer} &
{\centering  $N_{\rm H}$  } &
{\centering  $\Gamma$} &
{\centering  $kT_{\rm e}$} &
{\centering  $kT_{\rm BB}$} &
{\centering  red-$\chi^{2}$(d.o.f)}&
{\centering  $F_{\rm per}$$^{\mathrm{a}}$} &\\
(Obs Id)& $(10^{22}\enspace  \mathrm{cm^{-2}})$ & & $\mathrm{(keV)}$ & $\mathrm{(keV)}$   &  & ($10^{-9}\enspace \mathrm{erg\enspace s^{-1}\enspace cm^{-2}}$) \\ [0.01cm] \hline
4202200125  &                 & $1.91\pm{0.01}$&$2.19\pm{0.26}$  & $0.36\pm{0.01}$ & 0.99(120)&         $2.82\pm{0.02}$ \\
5202200101  &                 & $1.83\pm{0.01}$&$1.89\pm{0.15}$  & $0.34\pm{0.01}$ & 1.01(115)&         $2.87\pm{0.02}$ \\
5202200112  &                 & $1.83\pm{0.02}$&$1.90\pm{0.13}$  & $0.35\pm{0.01}$ & 1.03(122)&         $3.59\pm{0.03}$ \\
5202200113  &                 & $1.89\pm{0.05}$&$1.88\pm{0.15}$  & $0.35\pm{0.01}$ & 0.95(119)&         $3.32\pm{0.03}$ \\
4639010102  &                 & $2.06\pm{0.02}$&$2.64_{-0.43}^{+0.98}$  & $0.36\pm{0.01}$ & 1.08(118)&  $3.10\pm{0.03}$ \\
4639010104  &                 & $1.79\pm{0.02}$&$1.77\pm{0.11}$  & $0.34\pm{0.01}$ & 0.97(121)&         $3.33\pm{0.03}$ \\
4639010113  &                 & $1.84\pm{0.04}$&$1.87\pm{0.14}$  & $0.34\pm{0.01}$ & 1.18(120)&         $3.14\pm{0.03}$ \\
4639010116  &                 & $1.88\pm{0.05}$&$2.02\pm{0.19}$  & $0.35\pm{0.01}$ & 0.83(118)&         $2.96\pm{0.02}$ \\
4639010131  &0.413$\pm{0.001}$& $2.06\pm{0.02}$&$2.19_{-0.26}^{+0.42}$   & $0.35\pm{0.01}$ & 1.10(117)& $2.42\pm{0.02}$ \\ 
4639010141  &                 & $1.90\pm{0.02}$&$2.12\pm{0.23}$  & $0.36\pm{0.01}$ & 1.18(120)&         $2.70\pm{0.02}$ \\ 
4639010146  &                 & $1.98\pm{0.01}$&$1.83\pm{0.18}$  & $0.35\pm{0.01}$ & 1.12(117)&         $2.55\pm{0.02}$ \\ 
4639010160  &                 & $2.02\pm{0.17}$&$2.33_{-0.31}^{+0.57}$  & $0.35\pm{0.01}$ & 0.94(117)&  $2.20\pm{0.02}$ \\ 
4639010160  &                 & $2.21\pm{0.01}$&$3.04_{-0.80}^{+13.65}$ & $0.36\pm{0.01}$ & 1.00(109)&  $1.91\pm{0.01}$ \\ 
4639010166  &                 & $1.97\pm{0.01}$&$2.32_{-0.28}^{+0.45}$  & $0.34\pm{0.01}$ & 1.12(118)&  $2.20\pm{0.01}$ \\ 
4639010175  &                 & $2.19\pm{0.01}$&$2.67_{-0.60}^{+2.40}$  & $0.33\pm{0.02}$ & 0.84(110) & $1.31\pm{0.01}$ \\ 
4639010177  &                 & $2.20\pm{0.01}$&$2.05_{-0.25}^{+0.35}$ & $0.34\pm{0.02}$ & 0.99(102) & $1.07\pm{0.01}$ \\ 
4639010179  &                 & $2.10\pm{0.02}$&$2.36_{-0.41}^{+0.94}$  & $0.33\pm{0.02}$ & 1.14(109) & $1.27\pm{0.01}$ \\ 
\hline

\hline
\end{tabular} }

\end{center}

$^{\mathrm{a}}$ The unabsorbed bolometric persistent flux in $0.1-250$ keV.

\end{table*}

\subsection{X-ray burst time-resolved spectroscopy}
\label{Sec:spec_burst}

To investigate the time-resolved spectroscopy during X-ray bursts, we extracted the burst spectra with the exposure time varying between 0.125--4 s, to guarantee each spectrum had at least 2000 counts in the energy range between 0.3--10 keV. All burst spectra were grouped using {\tt grappha} with a minimum count of 20. To fit the burst spectra, we first used an absorbed blackbody model, \texttt{TBabs} $\times$ \texttt{bbodyrad}, where we regarded the persistent spectra being the background and remaining unchanged during bursts. The \texttt{bbodyrad} model has two parameters, the blackbody temperature, $T_{\rm bb}$, and the normalization, $K$. We fixed the hydrogen column density at $0.413 \times 10^{22}~{\rm cm} ^{-2}$.  We found that the blackbody model can fit the spectra well during the cooling tail. However, for the first $\sim$ 10 s of all 17 X-ray bursts, the model could not fit the spectra below 1.5 keV and above 5 keV. The red $\chi^{2}$ are higher than unity (see the gray dashed lines in Figs.~\ref{Fig:fa}  and \ref{Fig:fa2}). The fit can be significantly improved by using the enhanced persistent emission model (see Sec. \ref{Sec:spec_fa}) or by adding a reflection component from the surrounding accretion disk (see Sec.~\ref{Sec:spec_diskref}). %
   
\subsubsection{Enhanced persistent emission}
\label{Sec:spec_fa}
\begin{figure*}
    \centering
        \includegraphics[width=\hsize]{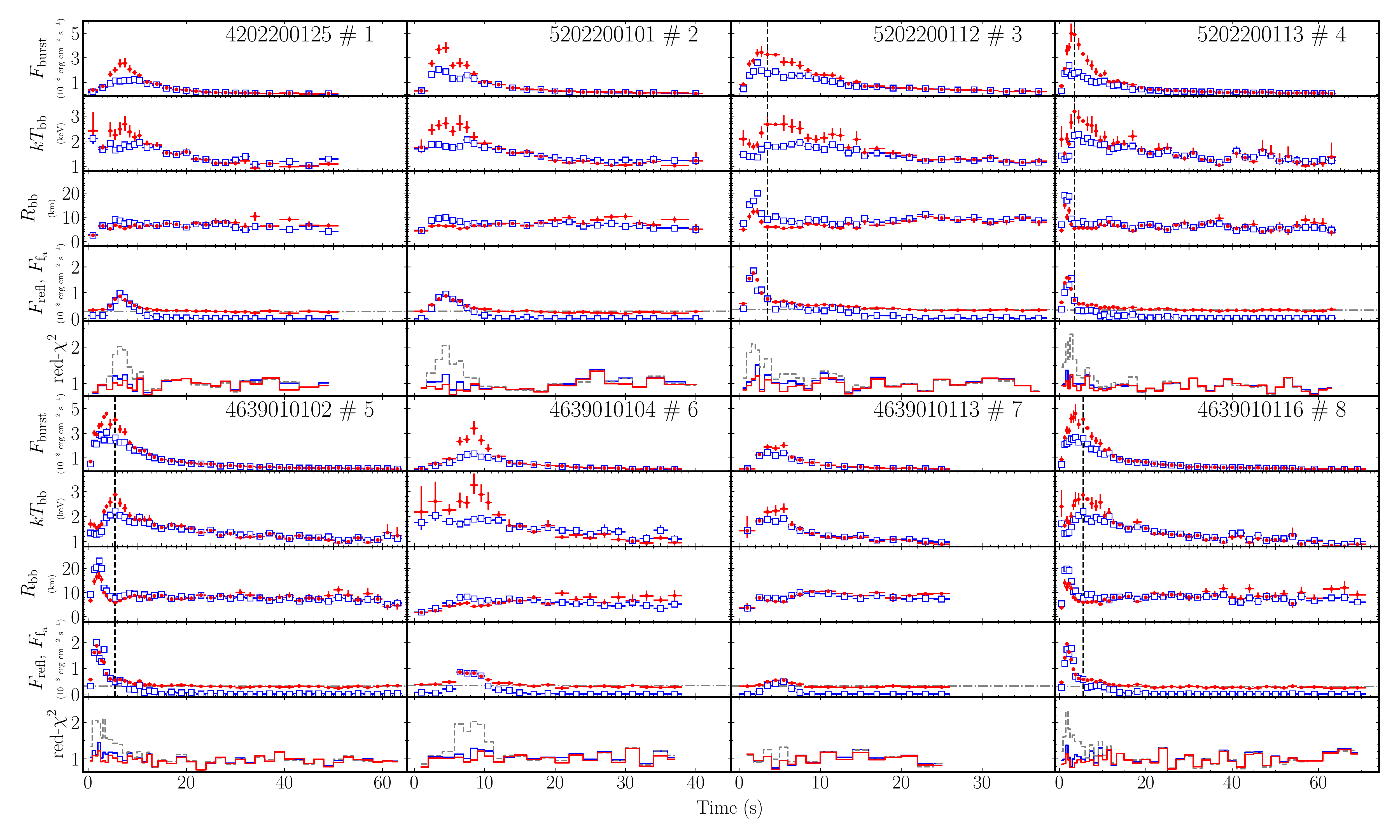}
        \caption{Time-resolved spectroscopy using the $f_a$ model (red dot) and the disk reflection model (blue square) for bursts \#1-8. For each panel, from top to bottom, we exhibit the burst bolometric flux, $F_\mathrm{ burst}$; the blackbody temperature, $kT_\mathrm{bb}$,  and the blackbody radii, $R_{\rm bb}$, which were calculated using a distance of 7.54 kpc; the enhanced persistent emission flux and disk reflection flux, $F_{\rm f_{a}}$ and $F_{\rm refl}$, respectively; and the goodness of fit per degree of freedom, red $\chi^{2}$.  The gray dashed-dotted lines mean the persistent emission level, and the black dashed lines label the touchdown moment of the $f_{\rm a}$ model. The red $\chi^2$ of the blackbody model are plotted as a gray dashed line for comparison.}            %
        \label{Fig:fa}%
    \end{figure*}

\begin{figure*}
    \centering
        \includegraphics[width=\hsize]{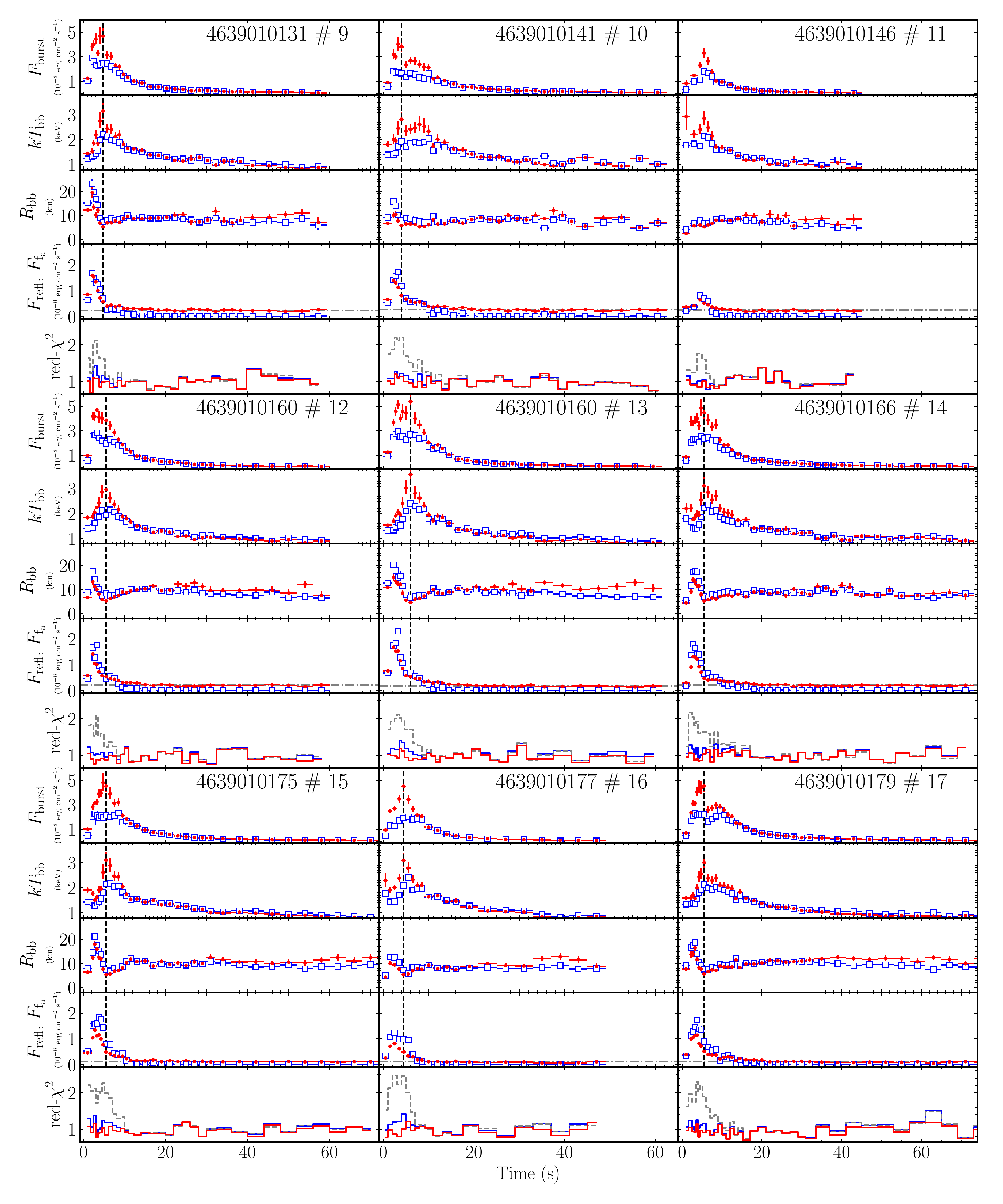}
        \caption{Similar to Fig.~\ref{Fig:fa}, but for bursts \#9-17.}
        \label{Fig:fa2}%
    \end{figure*}

To account for the residuals of the blackbody burst fit, we adopted the following  best-fit $f_{\rm a}$ model composed of \texttt{TBabs} $\times$ \texttt{(bbodyrad} +  $f_a\times$ \texttt{nthcomp)}. The ${\tt nthcomp}$ accounts for the persistent emission during the burst, and its parameters are fixed to the best-fitted values listed in Table \ref{table:preburst}. We assumed that the amplitude of the persistent emission can change during bursts. In this case, only the instrumental background was subtracted.  The parameter $f_{\rm a}$ is a free scaling factor that can be used to account for the persistent emission variation. If $f_{\rm a}=1$, this means that the amplitude of persistent emission during the burst is equal to the value before the burst. We show the best-fit values of all bursts by using the $f_{\rm a}$ method in Fig.~\ref{Fig:fa} and Fig.~\ref{Fig:fa2}, where the red $\chi^2$ of the blackbody model is plotted for comparison. For all bursts, the $f_{\rm a}$ values rise at the burst start and quickly reach the maximum within a few seconds; then they decrease slowly and return to around unity during the cooling tail. This indicates that during the expansion phase, due to the X-ray burst radiation, the persistent emissions rapidly enhanced, and then returned to the preburst level during the cooling tail.
Based on the time-resolved spectra, we found  that 12 out of 17 X-ray bursts showed significant PRE \citep{Galloway08}. For these 12 bursts, the blackbody radii expanded at the burst start and reached the maximum at around 20 km. When the atmosphere moved back to the NS surface, the blackbody temperature increased to its peak and the blackbody radii decreased to its local minimum which corresponds to  the touchdown moment. Afterwards, the atmosphere temperature decreased and the blackbody radii were around 10 km, implying that the whole NS surface cooled down.  

We measured the bolometric flux at the touchdown moment for all 12 PRE bursts.  We obtained the burst fluence, $E_{\rm b},$ by summing the measured flux  over the burst. The peak flux, $F_{\rm peak}$, the bolometric flux at the touchdown moment, $F_{\rm TD}$, and the burst fluence of each burst are listed in Table \ref{table:burst_ob}. 

\begin{table*}
\begin{center} 
\caption{Burst parameters' overview.  \label{table:burst_ob}}

\resizebox{\linewidth}{!}{\begin{tabular}{ccccccccccccccc} 
{\centering  Burst$^{\mathrm{a}}$  } &
{\centering  ObsId$^{\mathrm{b}}$ } &
{\centering  Burst Onset} &
{\centering  Peak Rate$^{\rm c}$} &
{\centering  $F_{\rm peak}^{\rm d}$  } &
{\centering  $E_{\rm b}$} &
{\centering  PRE  } &
{\centering  $F_{\rm TD}^{\rm e}$ } &
{\centering  $kT_{\rm TD}$ } &
{\centering  $\Delta T_{\mathrm{rec}}^{\rm f}$ } &
{\centering  $\Delta t_{\mathrm{rise}}^{\rm g}$ } &
{\centering  $\tau^{\rm h}$ } &
{\centering  $\alpha^{\rm i}$ } &
\\
 $\#$ & & (MJD) &($\mathrm{10^{3} ~ c ~ s^{-1}}$) &  &($10^{-7}~ \mathrm{erg~cm^{-2}}$)&  &  & (keV) &(hr) &(s)&(s) &\\ [0.01cm] \hline
-  & - & 59404.30775 & -    & - &-&Y&-&-&  -  &-&-&-&\\
1  &x25& 59404.55779 & 2.54 & $2.56\pm{0.38}$&$2.77\pm{0.32}$&N&       -       &       -       & 6.0   &4.6& $10.8\pm{2.0}$ &$235\pm{28}$&\\
-  & - & 59411.72027 & -    & - &-&Y&-&-&  171.9  &-&-&-&\\
2  &y01& 59639.33490 & 3.04 & $3.80\pm{0.44}$ & $3.10\pm{0.32}$ &N&       -       &       -       & -     &2.4& $ 8.2\pm{1.3}$ &-           &\\
3  &y12$^{\rm j}$& 59657.91325 & 5.53 & $3.46\pm{0.58}$&$4.14\pm{0.44}$&Y &$3.28\pm{0.41}$&$2.67\pm{0.25}$& 445.9 &0.9& $12.0\pm{2.4}$ &-           &\\
4  &y13 & 59658.96140 & 5.53 & $4.99\pm{0.73}$&$4.91\pm{0.57}$&Y &$4.89\pm{0.61}$&$3.18\pm{0.32}$& 25.2  &1.0& $ 9.8\pm{1.8}$ &-           &\\
5  &z02 & 59664.12261 & 6.78 & $4.60\pm{0.57}$ & $5.18\pm{0.43}$ & Y &$4.10\pm{0.49}$&$2.87\pm{0.25}$& 123.9 &1.0& $11.3\pm{1.7}$ &-           &\\
6  &z04 & 59666.95133 & 2.24 & $3.40\pm{0.54}$ & $2.46\pm{0.35}$ & N &       -       &       -       & 67.9  &3.6& $ 7.2\pm{1.5}$ &-           &\\
7  &z13 & 59675.59720 & 1.62 & $2.03\pm{0.23}$ & $1.39\pm{0.14}$ & N &       -       &       -       & 207.5 &1.1& $ 6.8\pm{1.0}$ &-           &\\
8  &z16 & 59678.77330 & 6.34 & $4.61\pm{0.62}$ & $5.23\pm{0.52}$ & Y &$4.13\pm{0.49}$&$2.86\pm{0.26}$& 76.2  &0.8& $11.3\pm{1.9}$ &-           &\\
9  &z31 & 59695.09695 & 7.35 & $4.68\pm{0.62}$ & $4.37\pm{0.40}$ & Y &$4.68\pm{0.62}$&$3.15\pm{0.34}$& 391.8 &0.9& $ 9.3\pm{1.5}$ &-           &\\
-  & -  & 59699.26102 & -    & - &-&Y&-&-&  99.9  &-&-&-&\\
-  & -  & 59702.29222 & -    & - &-&Y&-&-&  72.7  &-&-&-&\\
-  & -  & 59704.34823 & -    & - &-&Y&-&-&  49.3  &-&-&-&\\
-  & -  & 59705.15495 & -    & - &-&Y&-&-&  19.4  &-&-&-&\\
-  & -  & 59706.25532 & -    & - &-&Y&-&-&  26.4  &-&-&-&\\
-  & -  & 59707.34781 & -    & - &-&Y&-&-&  26.2  &-&-&-&\\
-  & -  & 59708.45073 & -    & - &-&N&-&-&  26.5  &-&-&-&\\
-  & -  & 59709.33130 & -    & - &-&Y&-&-&  21.1  &-&-&-&\\
10 &z41 & 59718.33075 & 4.87 & $4.03\pm{0.57}$ & $4.23\pm{0.45}$ & Y &$3.82\pm{0.44}$&$2.82\pm{0.24}$& 216.0 &0.5& $10.5\pm{1.9}$ &-           &\\
11 &z46 & 59723.82507 & 2.16 & $3.29\pm{0.42}$ & $2.49\pm{0.27}$ & N &       -       &       -       & 131.9 &4.0& $ 7.6\pm{1.3}$ &-           &\\
12 &z60 & 59739.42887 & 5.28 & $4.70\pm{0.16}$ & $4.50\pm{0.35}$ & Y &$3.86\pm{0.14}$&$2.97\pm{0.06}$& 374.5 &0.5& $ 9.6\pm{0.8}$ &-           &\\
13 &z60 & 59739.87450 & 6.42 & $5.40\pm{0.75}$ & $5.45\pm{0.48}$ & Y &$5.40\pm{0.75}$&$3.56\pm{0.40}$& 10.7  &0.8& $10.1\pm{1.7}$ &$127\pm{10}$&\\
14 &z66 & 59747.68374 & 5.71 & $4.84\pm{0.67}$ & $5.35\pm{0.52}$ & Y &$4.49\pm{0.57}$&$3.12\pm{0.30}$& 187.4 &1.3& $11.1\pm{1.9}$ &-           &\\
15 &z75 & 59756.86069 & 5.50 & $4.81\pm{0.79}$ & $5.13\pm{0.42}$ & Y &$4.53\pm{0.55}$&$3.10\pm{0.28}$& 220.2 &1.5& $10.7\pm{2.0}$ &-           &\\
$16^{\rm h}$ &z77 & 59760.00991 & 3.66 &  $4.51\pm{0.56}$ & $ 3.62\pm{0.31}$ &Y&$4.51\pm{0.56}$&$3.10\pm{0.28}$&75.6 &1.2& $8.0\pm{1.2}$ &-           &\\
17 &z79 & 59762.73725 & 5.51 & $4.54\pm{0.56}$ & $5.59\pm{0.44}$ & Y &$4.54\pm{0.56}$&$3.01\pm{0.28}$& 65.5 &1.8& $12.3\pm{1.8}$ &-           &\\
\hline  
\end{tabular} }

\end{center}
$^{\rm a}$ The X-ray bursts observed by \hxmt\ are  listed without numbers.

$^{\rm b}$ We only use the last two digits to represent \nicer\ ObsIDs, so x=42022001, y=52022001, and z=46390101. 

$^{\rm c}$ The peak rates with persistent emission being subtracted were measured from the 1 s light curves in the energy range of 0.5--10 keV.

$^{\rm d}$ The bolometric peak flux of each burst is in units of $10^{-8}~ \mathrm{erg~s^{-1} cm^{-2}}$. 

$^{\rm e}$ The bolometric touchdown flux is in units of $10^{-8}~ \mathrm{erg~s^{-1} cm^{-2}}$.

$^{\rm f}$ The observed recurrence time.

$^{\rm g}$ The time of burst onset to its peak.

$^{\rm h}$ The decay time of bursts defined as the ratio of the burst fluence to its peak flux.

$^{\rm i}$ The ratio of the integrated persistent flux to the burst fluence, respectively. 

$^{\rm j}$ The tail of burst \#3 was truncated due to a data gap.

$^{\rm h}$ Burst \#16 was observed in the UFA file.\\

\end{table*}


\subsubsection{The disk reflection model}
\label{Sec:spec_diskref}
We can also explain the deviations of the blackbody model to the burst spectra by adding a  disk reflection component. We used  the advanced  model  \texttt{relxillNS} to account for the disk reflection, which calculates a photoionized accretion disk illuminated by blackbody spectra radiated from a NS \citep{Garcia21}. The \texttt{relxillNS} model has been used to fit the continuum of a black hole and NSs in LMXBs \citep{Ludlam19,Connors20}, and also to explain the contribution from the accretion disk reflecting the burst radiation in 4U 1636--536 \citep{Zhao22}. We used the model \texttt{TBabs} $\times$ \texttt{(bbodyrad + nthcomp + relxillNS)} to fit the spectrum of the burst, where the \texttt{blackbody}, \texttt{nthcomp}, and \texttt{relxillNS} account for the X-ray burst emission, the persistent emission, and the reflection of the burst photons from the accretion disk, respectively. The parameters of \texttt{nthcomp} were fixed to the best-fit values reported in Table \ref{table:preburst}. The parameters of the \texttt{relxillNS} model are difficult to constrain simultaneously because the exposure of each burst spectrum is very short.  Only the normalization of \texttt{relxillNS} is free to change. We fixed the indices of emissivity for the coronal flavor models, $q_1 = q_2 = 3$, and the break radius between these two emissivity indices, $R_{\rm br}=15$. Assuming canonical NS mass $M_{\rm NS}=1.4M_{\odot}$ and radius $R_{\rm NS}=10$ km, as well as the observed spin frequency of 584.65 Hz \citep{Li22}, we obtained the dimensionless spin parameter, $a=0.27$. We assumed an inclination angle from the system to the observer of $i=30^{\circ}$, the inner and outer radii of the accretion disk, $R_{\rm in}=R_{\rm ISCO}$ and $R_{\rm out}=400R_{g}$, where $R_{\rm ISCO}$ is  the innermost stable circular orbit (ISCO) and $R_g = GM_{\rm NS}/c^2$ is the gravitational
radius, $G$ is the gravitational constant, and $c$ is the speed of light. We fixed the ionization parameter of the accretion disk at ${\rm log}\xi=3.2$, the solar iron abundance, $A_{\rm Fe}=4.5$, and the logarithmic value of the density, ${\rm log}n=10^{16}~\mathrm{cm}^{-3}$. 
The temperature of input spectrum $kT_{\rm BB,~seed}$ is tied with the  temperature of \texttt{bbodyrad}. We set the reflection fraction parameter $f_{\rm refl}=-1$, which means only the reflection emission was returned.

The fitted parameters of the disk reflection model are shown in Fig.~\ref{Fig:fa} and Fig.~\ref{Fig:fa2} and marked as blue squares, where $F_\mathrm{{refl}}$ is the flux of the \texttt{relxillNS} component calculated by using \texttt{cflux} in the energy range of 0.1--250 keV. We found that the disk reflection model can also model the derivations properly for all red $\chi^2\sim1.0$.  Compared with the results of the enhanced persistent emission model, the disk reflection model provides similar blackbody radii but lower temperatures, and thus lower bolometric fluxes.  The fluxes of the reflection component are statistically significant for the first $\sim10$ s for all bursts, and they can be neglected during the cooling tail.   We find the temperatures and bolometric flux of the disk reflection model to be less than the $f_\mathrm{{a}}$ model by a factor $\sim0.25$ and $2.3$, respectively, which is similar to \cite{Zhao22}. We also tried to coadd the spectra of all PRE bursts during the expansion phase to find spectral features, but we detected none. 

\section{Discussion}
\label{Sec:discussion}

We analyzed the 2021 and 2022 outbursts of 4U 1730--22 from \nicer\ observations. These two outbursts showed a typical behavior of atoll sources and  similar spectral evolution from the HID. We carried out detailed time-resolved spectral analyses on all 17 type I X-ray bursts. Each persistent spectrum prior to the trigger of these bursts could be well fitted by an absorbed thermally Comptonized model in the energy range 0.3--10 keV. We found that the burst spectra for the first $\sim10$ s after onset were deviated from the blackbody model. Therefore, we introduced the  $f_a$ model and the reflection model to account for it. Both the $f_a$ model and the reflection model explain the burst spectra well \citep[see also][]{Zhao22}, and they are difficult to distinguish by using current on-orbit instruments. We found that the contribution of the enhanced persistent emission or the disk reflection of PRE bursts are more significant than non-PRE bursts. This is not surprising because PRE bursts with $E_{\rm b}\sim3.3-5.4\times 10^{-7}~{\rm erg~cm^{-2}}$ are more energetic than non-PRE bursts with $E_{\rm b}\sim1.4-3.1\times 10^{-7}~{\rm erg~cm^{-2}}$, and thus they can induce more significant enhanced persistent emission or disk reflection. The touchdown fluxes determined from the reflection model are lower than the results from the $f_a$ model. However, when the burst photons are reflected, the observed burst flux could be blocked and/or strengthened by the accretion disk. The anisotropy of the burst emission depends on the inclination angle of the disk to the line of sight and the shape of the disk \citep{He16}. For this work, we have used the results from the $f_a$ model to estimate the distance to 4U 1730--22 (Sec.~\ref{sec:s_distance}) and to determine the burning fuel from the recurrence time (Sec.~\ref{sec:fuel}). %

From the time-resolved spectra, 12 out of 17 bursts are PRE bursts; when accounting for nine PRE bursts out of ten X-ray bursts observed by \hxmt, 78\% of X-ray bursts from 4U 1730--22 exhibited PRE. Such a large fraction of PRE bursts have also been found in 4U 1820--30, 4U 1916--053, and 4U 1728--34, which was the case for the ignition of pure helium in a hydrogen-poor environment \citep{Galloway08,Galloway20}. The light-curve profiles of all 12 PRE bursts from 4U 1730--22 are similar, and they have the rise time of 0.5--2 s and the decay time, $\tau=E_{\rm b}/F_{\rm peak} $,  of $\sim 10$ s. The observed rise and decay times are close to the case of helium bursts from 4U 1820--30, 4U 1916--053, and 4U 1728--34  with the decay time $\tau\sim5-9$ s \citep{Galloway08,Galloway20}, but they are smaller than the mixed hydrogen and helium bursts, such as XTE J1739--285, with the rise time of 3.6--11.5 s and the decay time of $\sim 20$ s \citep{2021ApJ_Bult}. The unstable burning of helium through the triple-alpha process is usually faster than the unstable burning of mixed hydrogen and helium. Thus, we propose that these PRE bursts are powered by the unstable burning of pure helium, following the completion of hydrogen burning, since the companion of 4U 1730--22 is likely in the main sequence \citep{Strohmayer06,ATel14693,ATel14694}. 

\subsection{The source distance}
\label{sec:s_distance}

Assuming spherically symmetric emission, we can estimate the Eddington luminosity measured by the observer at infinity as \citep{Lewin93, Suleimanov12,Poutanen17}
\begin{equation}
\begin{split}
L_{\rm Edd,~\infty }
&=\frac{8\pi Gm_{p} M_{\rm NS}c[1+(kT/39.4~\rm{keV})^{0.976}]}{\sigma_{\rm T}(1+X)(1+z)}  \\
&=2.7\times 10^{38}\Bigr (\frac{M_{\rm NS}}{1.4M_\odot }\Bigr )\frac{1+(kT/39.4~{\rm keV})^{0.976} }{(1+X)}\\
&\quad \times \Bigl (\frac{1+z}{1.31}\Bigr)^{-1}~\mathrm{erg~s^{-1}}, 
\end{split}
\label{eq:L_edd} 
\end{equation}
where $m_{\rm p}$ is the mass of the proton, $\sigma_{\rm T}$ is the Thompson scattering cross section, $kT$ is the effective temperature of the atmosphere in units of keV, the gravitational redshift factor $1+z=(1-2GM_{\rm NS}/Rc^2)^{-1/2}$ at the radius of $R$, the mass fraction of hydrogen $X=0$ for pure helium, and $X=0.7$ for solar abundance. We assumed the blackbody temperature at the touchdown moment as the effective temperature, that is $kT=kT_{\rm TD}$. For a typical NS with a mass of $1.4M_{\odot}$ and the atmosphere being located on the NS surface at the radius of 10 km, we obtained the Eddington luminosity of pure helium and the solar abundance, $L_{\rm Edd,~\infty }\sim2.93\times10^{38}~{\rm erg~s^{-1}}$ and $1.72\times10^{38}~{\rm erg~s^{-1}}$, respectively.  We calculated the distance by using the touchdown flux via the relation $D=(L_\mathrm{Edd,~\infty}/4\pi F_\mathrm{TD,~PRE})^{1/2}$, then we obtained the average distance $ d=7.54\pm{0.46}\;kpc$ for $X=0$, and $d=5.79\pm{0.35}$ kpc for $X=0.7$. We note that the error of each observable parameter was propagated to estimate the uncertainties of the distance and other quantities in Table~\ref{table:burst_ob} and \ref{table:Calculated}. 
The distance is smaller than the result from \citet{2022arXiv220910721C}, that is 10.4 kpc, in which they used the empirical value $3.8\times10^{38}~{\rm erg~s^{-1}}$ as the Eddington luminosity \citep{Kuulkers03} and a smaller touchdown flux, $F_{\rm TD}=2.92\pm0.11\times10^{-8}~{\rm erg~cm^{-2}~s^{-1}}$, measured by \hxmt\ possibly due to the adopted different energy bands between \hxmt\ and \nicer. If we take the same Eddington luminosity,  $3.8\times10^{38}~{\rm erg~s^{-1}}$, as was used in Chen et al. (2022), the distance is $8.59\pm{0.54}$ kpc closer to the distance value we derived using Eq. \ref{eq:L_edd}.  Hereafter, we use 7.54 kpc as the distance to the source.

\subsection{X-ray burst fuel}
\label{sec:fuel}
Besides the rise and decay time of burst light curves, we were able to determine the burst fuel in an independent way. We calculated the $\alpha$ factor to verify the burst fuel composition, which is the ratio of the persistent fluence between bursts to the total burst fluence of the burst,  $\alpha=F_\mathrm{per} \Delta T_\mathrm{rec}/E_\mathrm{b}$.
We obtained the $\alpha$ of the burst with $\Delta T_{\mathrm{rec}}<1$ day, and listed all $\alpha$ in Table \ref{table:burst_ob}. The calculated $\alpha$ is the upper limit, which was hard to verify for the burst fuel composition. Instead, we estimated the averaged recurrence time by dividing the unfiltered exposure of 467.8 ks collected between the first and last bursts from ObsIDs  5202200101 and 4639010179 during the 2022 outburst, respectively, by the number of bursts observed. We obtained the mean recurrence time $\bar{\Delta T_{\mathrm{rec}}} \approx  8.1$ hours. From the average values of burst fluence, persistent flux, and $\bar{\Delta T_{\mathrm{rec}}}$, we found the average $\Bar{\alpha} \approx  173$, which is close to a helium fuel composition. Even if eight X-ray bursts are added during the 2022 outburst from \hxmt\  without an extra exposure time, $\Bar{\alpha}\approx115$ is still larger than the mixed hydrogen and helium fuel.

The burst fuel composition can be verified in another way by determining the local accretion rate \citep{Galloway08}, 
\begin{equation}
\begin{split}
    \Dot{m}
    &=\frac{L_\mathrm{ {per}}(1+z)}{ 4\pi R_{\rm NS}^{2}(GM_{\rm NS}/R_{\rm NS})}\\
    &\approx 6.7\times 10^{3}\biggl(\frac{F_\mathrm{{per}}}{10^{-9}\mathrm{~ergs ~ cm^{-2}~s^{-1}}}\biggr)\biggl(\frac{d}{10\mathrm{~ kpc}}\biggr)^{2}\biggl(\frac{M_{\rm NS}}{1.4M_{\odot}}\biggr)^{-1}\\
    &\quad\times\biggl(\frac{1+z}{1.31} \biggr)\biggl(\frac{R\mathrm{_{NS}}}{10\mathrm{~ km}}\biggr)^{-1}\mathrm{g~cm^{-2}}\mathrm{~s^{-1}},
    \label{eq:lo_accration}
\end{split}      
\end{equation}\\ 
where $F_{\rm per}$ is the persistent flux.  
We used distance $ d=7.54\pm{0.46}\;kpc$ to calculate  $\Dot{m}$. We assumed the local Eddington accretion rate $\Dot{m}_{\rm Edd}=8.8\times10^{4} \mathrm{\frac{1.7}{(X+1)}~g~cm}^{-2} \mathrm{~s}^{-1}$, then we obtained the local accretion rate, $\Dot{m}$, as a fraction of the local Eddington accretion rate, $\Dot{m}_{\rm Edd}$, for $X=0.7$ and $X=0$, respectively. We listed $\Dot{m}$ and $\Dot{m}/\Dot{m}_{\rm Edd}$ in Table \ref{table:Calculated}. All $\Dot{m}$  are less than 10\% $\Dot{m}_{\rm Edd}$ for $X=0$, which supports the idea that these bursts were ignited in a  helium-rich environment \citep[see e.g.,][]{Galloway21}. For the case of $X=0.7$,  most $\Dot{m}$ were higher than 10\% $\Dot{m}_{\rm Edd}$, but for bursts \#15-17, $\Dot{m}\sim6\%~\Dot{m}_{\rm Edd}$, which does not support the mixed hydrogen and helium bursts.

We can estimate the ignition depth at the onset of the burst with the equation 
\begin{equation}
y_{\rm ign}=\frac{4\pi E_\mathrm{{b}}d^{2}(1+z)}{4\pi R_{\rm NS}^{2}Q\mathrm{_{nuc}}}, \label{eq:ign}
\end{equation} 
where  $E_\mathrm{{b}}$ is the burst fluence, the nuclear energy generated for solar composition ($X=0.7$) is $Q_\mathrm{{nuc}} \approx 1.31+6.95X -1.92X^{2}\mathrm{~MeV~nucleon^{-1}}\approx 4.98 \mathrm{~MeV~nucleon^{-1}}$  and $Q_\mathrm{{nuc}} \approx 1.31 \mathrm{~MeV~nucleon^{-1}}$ for $X=0$  \citep{Goodwin19}. We used $ d=7.54\pm{0.46}\;kpc$ to determine  $y_\mathrm{ign}$.
Once the ignition depth was known, we calculated the recurrence time between bursts by using the equation $\Delta t_{\mathrm{rec}}=(y_{\mathrm{ign}} / \Dot{m})(1+z)$. The estimated recurrence time and ignition depth are listed in Table \ref{table:Calculated}.
The recurrence times for $X=0$ are four times longer than $X=0.7$.%

The $\Delta t_\mathrm{{rec}}$ for $X=0$ between the first \hxmt\ burst and burst \#1 is $5.40\pm{1.18}$ hr, which is consistent with the observed recurrence time $\Delta T_{\mathrm{rec}}=6.0$ hr. The calculated recurrence time $\Delta t_\mathrm{{rec}}$ between bursts \#12 and \#13 is $15.67\pm{3.10}$ hr, which is slightly higher than the observed result $\Delta T_{\mathrm{rec}} =10.7~\mathrm{hr}$. The  reason is that the local accretion rate ($\Dot{m}$) decreased between these two bursts.
Other calculated recurrence times are prominently less than the observed recurrence time, which means that some bursts were likely missed due to data gaps.  We verified that at the predicted burst recurrence time, we always had an observational data gap for $X=0$, but not for  $X=0.7$.   
We also computed the average recurrence time, and found $\bar{\Delta t_\mathrm{{rec}}} =  2.73\pm{0.54}~\mathrm{and}~10.88\pm{2.18}$ hr for $X=0.7$ and 0, respectively. We note that due to burst \#1 appearing in the 2021 outburst, all of the calculated average values do not include it.  For $X=0$, the $\bar{\Delta t_{\rm rec}}$ within the error range is close to the observed mean recurrence time $\bar{\Delta T_{\rm rec}}=8.1$ hours. For $X=0.7$, the predicted recurrence time $\bar{\Delta t_{\rm rec}}$ is prominently less than $\bar{\Delta T_{\rm rec}}$. If these bursts were powered by mixed hydrogen and helium, we would expect to observe more X-ray bursts within 467.8 ks. Therefore, in regards to the magnitude, we conclude that these bursts from 4U~1730--22 occurred in a helium-rich environment, where helium was mainly produced from the steady burning of the accreted hydrogen from the main sequence companion. 

We note that \citet{2022ApJ...940...81B} also reported the analysis of type I X-ray bursts independently from the same data set of \nicer\ observations. Compared with their work, we found that the persistent spectra did not show the disk blackbody component significantly. We obtained a smaller hydrogen column density, $0.413\times10^{22}~{\rm cm^{-2}}$ versus $0.71\times10^{22}~{\rm cm^{-2}}$, and smaller red \chiq{}; however, our $N_{\rm H}$ value is consistent with the result report in \citet{Tomsick07}. The different persistent spectra and $N_{\rm H}$ can explain why  \citet{2022ApJ...940...81B} obtained larger peak fluxes and burst fluences by using the $f_a$ model. However, the distance to the source in our work is consistent with \citet{2022ApJ...940...81B} within a $1\sigma$ confidence level.

\begin{table*}
\begin{center} 
\caption{Calculated parameters.  \label{table:Calculated}}
\resizebox{\linewidth}{!}{\begin{tabular}{ccccccccccccc} 
 & &  &$X=0.$ & && & $X=0.7$ && & \\
\cline{3-5} \cline{7-9}\\
{\centering  Burst } &
{\centering  $\Dot{m}$} &
{\centering  $\Dot{m}/\Dot{m}_{\rm Edd}$} &
{\centering  $y_{\rm ign}$} &
{\centering  $\Delta t_\mathrm{{rec}}^{a}$ } & 
&
{\centering  $\Dot{m}/\Dot{m}_{\rm Edd}$} &
{\centering  $y_{\rm ign}$} &
{\centering  $\Delta t_\mathrm{{rec}}^{a}$ } & \\
 $\#$ & ($\mathrm{10^{4} ~ g ~ cm^{-2} ~ s^{-1}}$) &\% & ($10^{8}\mathrm{~g~ cm^{-2}}$) &(hr)&&\%&($10^{8}\mathrm{~g~ cm^{-2}}$) &(hr) &\\ [0.01cm] \hline
1&$1.08\pm{0.14}$&$7.2\pm{0.9}$&$1.60\pm{0.29}$&$5.40\pm{1.18}$& &$12.3\pm{1.6}$&$0.40\pm{0.07}$&$1.35\pm{0.30}$&\\
H1$^{b}$&$1.18\pm{0.15}$&$7.9\pm{1.0}$&$2.41\pm{0.22}$&$7.45\pm{1.18}$& &$13.4\pm{1.7}$&$0.61\pm{0.05}$&$1.87\pm{0.30}$&\\
2&$1.10\pm{0.14}$&$7.3\pm{0.9}$&$1.79\pm{0.30}$&$5.93\pm{1.24}$& &$12.5\pm{1.6}$&$0.45\pm{0.07}$&$1.48\pm{0.31}$&\\
3&$1.37\pm{0.17}$&$9.2\pm{1.2}$&$2.39\pm{0.41}$&$6.33\pm{1.34}$& &$15.6\pm{2.0}$&$0.60\pm{0.10}$&$1.59\pm{0.34}$&\\
4&$1.27\pm{0.16}$&$8.5\pm{1.1}$&$2.84\pm{0.51}$&$8.12\pm{1.79}$& &$14.4\pm{1.8}$&$0.71\pm{0.13}$&$2.03\pm{0.45}$&\\
5&$1.19\pm{0.15}$&$7.9\pm{1.0}$&$2.99\pm{0.44}$&$9.18\pm{1.78}$& &$13.5\pm{1.7}$&$0.75\pm{0.11}$&$2.30\pm{0.45}$&\\
6&$1.28\pm{0.16}$&$8.5\pm{1.1}$&$1.42\pm{0.29}$&$4.06\pm{0.98}$& &$14.5\pm{1.8}$&$0.36\pm{0.07}$&$1.02\pm{0.25}$&\\
7&$1.20\pm{0.15}$&$8.0\pm{1.0}$&$0.80\pm{0.13}$&$2.43\pm{0.51}$& &$13.7\pm{1.7}$&$0.20\pm{0.03}$&$0.61\pm{0.13}$&\\
8&$1.13\pm{0.14}$&$7.6\pm{1.0}$&$3.02\pm{0.49}$&$9.70\pm{2.01}$& &$12.9\pm{1.6}$&$0.76\pm{0.12}$&$2.43\pm{0.50}$&\\
9&$0.93\pm{0.12}$&$6.2\pm{0.8}$&$2.53\pm{0.39}$&$9.92\pm{1.99}$& &$10.5\pm{1.3}$&$0.63\pm{0.10}$&$2.48\pm{0.50}$&\\
H2$^{b}$&$1.18\pm{0.15}$&$7.9\pm{1.0}$&$1.66\pm{0.14}$&$5.12\pm{0.78}$& &$13.4\pm{1.7}$&$0.42\pm{0.04}$&$1.28\pm{0.20}$&\\
10&$1.03\pm{0.13}$&$6.9\pm{0.9}$&$2.44\pm{0.42}$&$8.60\pm{1.83}$& &$11.7\pm{1.5}$&$0.61\pm{0.10}$&$2.15\pm{0.46}$&\\
11&$0.98\pm{0.12}$&$6.5\pm{0.8}$&$1.44\pm{0.25}$&$5.36\pm{1.15}$& &$11.1\pm{1.4}$&$0.36\pm{0.06}$&$1.34\pm{0.29}$&\\
12&$0.84\pm{0.11}$&$5.6\pm{0.7}$&$2.60\pm{0.37}$&$11.23\pm{2.14}$& &$9.6\pm{1.2}$&$0.65\pm{0.09}$&$2.81\pm{0.53}$&\\
13&$0.73\pm{0.09}$&$4.9\pm{0.6}$&$3.15\pm{0.48}$&$15.67\pm{3.10}$& &$8.3\pm{1.1}$&$0.79\pm{0.12}$&$3.92\pm{0.78}$&\\
14&$0.84\pm{0.11}$&$5.6\pm{0.7}$&$3.09\pm{0.50}$&$13.36\pm{2.74}$& &$9.6\pm{1.2}$&$0.77\pm{0.12}$&$3.34\pm{0.68}$&\\
15&$0.50\pm{0.06}$&$3.4\pm{0.4}$&$2.97\pm{0.43}$&$21.51\pm{4.15}$& &$5.7\pm{0.7}$&$0.74\pm{0.11}$&$5.38\pm{1.04}$&\\
16&$0.41\pm{0.05}$&$2.7\pm{0.3}$&$2.08\pm{0.31}$&$18.48\pm{3.59}$& &$4.7\pm{0.6}$&$0.52\pm{0.08}$&$4.63\pm{0.90}$&\\
17&$0.49\pm{0.06}$&$3.3\pm{0.4}$&$3.23\pm{0.46}$&$24.18\pm{4.61}$& &$5.5\pm{0.7}$&$0.81\pm{0.11}$&$6.05\pm{1.15}$&\\
\hline  
\end{tabular}} 
\end{center}
$^{a}~\Delta t_\mathrm{{rec}}$ is the estimate for the recurrence time (see Sec.~\ref{sec:fuel} for more details)\\
$^{b}$ H1 and H2 represent the average data of two X-ray bursts observed by \hxmt\ in the 2021 outburst and eight X-ray bursts in the 2022 outburst, respectively.
\end{table*}

\section{Conclusions}
\label{Sec:conclusion}
We detected 17 type I X-ray bursts from 4U 1730--22 during its 2021 and 2022 outbursts by \nicer\ observations. The persistent spectra of all bursts are well fitted by \texttt{TBabs} $\times$ \texttt{nthcomp}, which showed a similar spectral shape. We have analyzed the time-resolved spectra of all bursts in detail. We found that the burst spectra in the first few seconds of onset are deviated from a blackbody shape below $\sim1.5$ keV and above $\sim5$ keV. The deviation could be explained by the enhanced persistent emission caused by the Poynting-Robertson drag \citep{Zand13} and the reflection of the burst emission from the surrounding accretion disk \citep[see e.g.,][]{Ballantyne04,Zhao22}. Specifically, we introduced the $f_{\rm a}$ model and the reflection model \texttt{relxillNS} to fit the time-resolved spectra of the burst, respectively. 
Due to the limited energy band of the observations, it is difficult to distinguish these two models. It is likely that the persistent emissions were enhanced and the accretion disk reflects part of X-ray burst photons at the same time during X-ray bursts.
We found that 12 of 17 type I X-ray bursts are PRE bursts. Accompanied with nine bursts out of ten being PRE bursts detected by \hxmt, 78\% of X-ray bursts from 4U 1730--22 showed PRE, systematically higher than the MINBAR catalog, that is, 20\% of PRE bursts. We propose that these PRE bursts were powered by pure helium based on the raise time, the recurrence time, and the local accretion rate.

\begin{acknowledgements}
We appreciate  the referee for valuable comments that improved this manuscript. Z.L. and Y.Y.P. were supported by National Natural Science Foundation of China (12130342, 12273030, U1938107). This work is supported by the National Key R\&D Program of China (2021YFA0718500), the National Natural Science Foundation of China under grants  11733009, U1838201, U1838202, U1938101, U2038101. This research has made use of data obtained from the High Energy Astrophysics Science Archive Research Center (HEASARC), provided by NASA’s Goddard Space Flight Center.
\end{acknowledgements}

\bibliography{name}
\bibliographystyle{aa}

\end{document}